\documentclass[a4paper,orivec]{llncs}
\usepackage{makeidx}  

\usepackage{endnotes}

\usepackage[utf8]{inputenc}
\usepackage[T1]{fontenc}

\usepackage{changepage} 

\usepackage[english]{babel}

\usepackage{url}

\newcommand{\hairspace}{\hspace{1pt}}
\newcommand{\eg}{\mbox{e.\hairspace{}g}.\ }  
\newcommand{\ie}{\mbox{i.\hairspace{}e}.\ }  

\newcommand{\etal}{\mbox{et~al.}\ }

\usepackage{mathpartir}

\usepackage[cmex10]{amsmath}
\usepackage{amssymb}

\usepackage{bbding}
\usepackage{tikz}

\usepackage{listings}
\usepackage{cite}

\usepackage{MnSymbol}

\usepackage[binary-units=true]{siunitx}

\usepackage{IEEEtrantools}

\mathchardef\mhyphen="2D 

\usepackage{booktabs}
\usepackage{array}
\usepackage{multirow}

\usepackage{pifont}
\usepackage{expdlist}

\usepackage{fancyvrb}

\usepackage[pdftex,hyperfootnotes=false,pdfpagelabels,bookmarks,linktoc=section]{hyperref}
\hypersetup{
    colorlinks=true,
    linktocpage=true,
    breaklinks=true,
    bookmarksnumbered,
    bookmarksopen=true,
    bookmarksopenlevel=1,
    pdfhighlight=/O,
    pdfpagemode=UseOutlines,
    urlcolor=black,
    linkcolor=black,
    citecolor=black,
    pdfauthor={Cornelius Diekmann}
}

\newcommand*\circled[1]{\tikz[baseline=-3pt, scale=0.5, every node/.style={scale=0.5}]{\node[shape=circle,draw,inner sep=1pt,minimum size=16pt] (char) {#1};}}

\newcommand\allow{\ensuremath{\text{\circled{\small \Checkmark}}}}
\newcommand\deny{\ensuremath{\text{\circled{\small \XSolidBrush}}}}
\newcommand\undecided{\ensuremath{\text{\circled{\textnormal{\large \textbf{?}}}}}}

\newcommand\matchop[1]{\ensuremath{\triangleright_{#1}}}
\newcommand\matches[1]{\ensuremath{#1 \,\matchop\gamma p}}
\newcommand\nmatches[1]{\ensuremath{\neg\; #1 \,\matchop\gamma p}}
\newcommand\bigstep[3]{\ensuremath{p \vdash\big\langle #1,\; #2 \big\rangle \Rightarrow #3}}

\newcommand\lstapp{\ensuremath{\mathbin{:\mkern-1mu:\mkern-1mu:}}} 
\newcommand\lstcons{\ensuremath{\mathbin{:\mkern-1mu:}}} 

\newcommand\bigstepapprox[4]{\ensuremath{p \vdash\big\langle #1,\; #2 \big\rangle \Rightarrow_{#4} #3}}

\newcommand\thmskip[0]{\vspace{-.1em}}

\newcommand\iptables{\emph{iptables}}

\usepackage{paralist}

\begin{document}

\title{Semantics-Preserving Simplification\\of Real-World Firewall Rule Sets}
\titlerunning{Semantics-Preserving Simplification of Firewall Rule Sets}

\author{Cornelius Diekmann
\and
Lars Hupel
\and
Georg Carle
}

\institute{Technische Universit{\"a}t M{\"u}nchen}%

\maketitle

\thispagestyle{empty}

\begin{abstract}
The security provided by a firewall for a computer network almost completely depends on the rules it enforces.
For over a decade, it has been a well-known and unsolved problem that the quality of many firewall rule sets is insufficient.
Therefore, there are many tools to analyze them.
However, we found that none of the available tools could handle typical, real-world \iptables{} rulesets.
This is due to the complex chain model used by \iptables{}, but also to the vast amount of possible match conditions that occur in real-world firewalls, many of which are not understood by academic and open source tools.

In this paper, we provide algorithms to transform firewall rulesets. 
We reduce the execution model to a simple list model and use ternary logic to abstract over all unknown match conditions.
These transformations enable existing tools to understand real-world firewall rules, which we demonstrate on four decently-sized rulesets.
Using the Isabelle theorem prover, we formally show that all our algorithms preserve the firewall's filtering behavior.
\end{abstract}

\begin{footnotesize}
\end{footnotesize}

\section{Introduction}
Firewalls are a fundamental security mechanism for computer networks.
Several firewall solutions, ranging from open source \cite{iptables,nftables,bsdpf} to commercial~\cite{ciscofirewallaccesslists,hpaclfirewalls}, exist.
Operating and managing firewalls is challenging as rulesets are usually written manually. 
While vulnerabilities in the firewall software itself are comparatively rare, it has been known for over a decade~\cite{firwallerr2004} that many firewalls enforce poorly written rulesets.
However, the prevalent methodology for configuring firewalls has not changed.
Consequently, studies regularly report insufficient quality of firewall rulesets \cite{fireman2006,netsecconflicts,ZhangAlShaer2007flip,databreach2009src,nelson2010margrave,sherry2012making,fwviz2012,diekmann2014verifying}.

Therefore, several tools~\cite{marmorstein2005itval,marmorstein2006firewall,fireman2006,tongaonkar2007inferring,cspfirewall,nelson2010margrave,fwviz2012,fwbuilder} have been developed to ease firewall management and reveal configuration errors.
However, when we tried to analyze real-world firewalls with the publicly available tools, none of them could handle our firewall rules. 
We found that the firewall model of the available tools is too simplistic.

In this paper, we address the following fundamental problem: Many tools do not understand real-world firewall rules. 
To solve the problem, we transform and simplify the rules such that they are understood by the respective tools.

\begin{figure*}[htb]
\begin{minipage}{\linewidth}
\footnotesize
\begin{Verbatim}[commandchars=\\\{\},codes={\catcode`$=3\catcode`^=7}]
Chain INPUT (policy ACCEPT)
target      prot source         destination  
DOS_PROTECT all  0.0.0.0/0      0.0.0.0/0  
ACCEPT      all  0.0.0.0/0      0.0.0.0/0   state RELATED,ESTABLISHED
DROP        tcp  0.0.0.0/0      0.0.0.0/0   tcp dpt:22
DROP        tcp  0.0.0.0/0      0.0.0.0/0   multiport dports $\hookleftarrow$
                                     21,873,5005,5006,80,548,111,2049,892
DROP        udp  0.0.0.0/0      0.0.0.0/0   multiport dports $\hookleftarrow$
                                                    123,111,2049,892,5353
ACCEPT      all  192.168.0.0/16 0.0.0.0/0  
DROP        all  0.0.0.0/0      0.0.0.0/0  

Chain DOS_PROTECT (1 references)
target      prot source         destination  
RETURN      icmp 0.0.0.0/0      0.0.0.0/0   icmptype 8 limit: $\hookleftarrow$
                                                        avg 1/sec burst 5
DROP        icmp 0.0.0.0/0      0.0.0.0/0   icmptype 8
RETURN      tcp  0.0.0.0/0      0.0.0.0/0   tcp flags:0x17/0x04 $\hookleftarrow$
                                                 limit: avg 1/sec burst 5
DROP        tcp  0.0.0.0/0      0.0.0.0/0   tcp flags:0x17/0x04
RETURN      tcp  0.0.0.0/0      0.0.0.0/0   tcp flags:0x17/0x02 $\hookleftarrow$
                                           limit: avg 10000/sec burst 100
DROP        tcp  0.0.0.0/0      0.0.0.0/0   tcp flags:0x17/0x02
\end{Verbatim}
\end{minipage}%
\caption{Linux \iptables{} ruleset of a Synology NAS (network attached storage) device}%
  \label{fig:firewall:synology}%
\end{figure*}

To demonstrate the problem by example, we decided to use \emph{ITVal}~\cite{marmorstein2005itval} because it natively supports \iptables{}~\cite{iptables}, is open source, and supports calls to user-defined chains. 
However, ITVal's firewall model is representative of the model used by the majority of tools; therefore, the problems described here also apply to a vast range of other tools.
Firewall models used in related work are surveyed in Sect.~\ref{sec:firewall-models}.
For this example, we use the firewall rules in Fig.~\ref{fig:firewall:synology}, taken from an NAS device.
The ruleset reads as follows: 
First, incoming packets are sent to the user-defined \texttt{DOS\_PROTECT} chain, where some rate limiting is applied. 
Afterwards, the firewall allows all packets which belong to already established connections. 
This is generally considered good practice.
Then, some services, identified by their ports, are blocked.
Finally, the firewall allows all packets from the local network 192.168.0.0/16 and discards all other packets.
We used ITVal to partition the IP space into equivalence classes (\ie ranges with the same access rights)~\cite{marmorstein2006firewall}.
The expected result is a set of two IP ranges: the local network 192.168.0.0/16 and the ``rest''.
However, ITVal erroneously only reports one IP range: the universe.
Removing the first two rules (in particular the call in the \texttt{DOS\_PROTECT} chain) lets ITVal compute the expected result.

We identified two main problems which prevent tools from ``understanding'' real-world firewalls. 
First, calling and returning from custom chains, due to the possibility of complex nested chain calls. 
Second, more seriously, most tools do not understand the firewall's match conditions.
In the above example, the rate limiting is not understood.
The problem of unknown match conditions cannot simply be solved by implementing the rate limiting feature for the respective tool.
The major reason is that the underlying algorithm might not be capable of dealing with this special case.
Additionally, firewalls, such as \iptables{}, support numerous match conditions and several new ones are added in every release.\endnote{As of version 1.4.21 (Linux kernel 3.13), \iptables{} supports more than 50 match conditions.} 
We expect even more match conditions for nftables~\cite{nftables} in the future since they can be written as simple userspace programs~\cite{whylovenftables2014blog}.
%
%
Therefore, it is virtually impossible to write a tool which understands all possible match conditions.

In this paper, we build a fundamental prerequisite to enable tool-supported analysis of \emph{real-world} firewalls: 
We present several steps of semantics-preserving ruleset simplification, which lead to a ruleset that is ``understandable'' to subsequent analysis tools:  
First, we unfold all calls to and returns from user-defined chains.
This process is exact and valid for arbitrary match conditions.
Afterwards, we process unknown match conditions.
For that, we embed a ternary-logic semantics into the firewall's semantics.
Due to ternary logic, all match conditions not understood by subsequent analysis tools can be treated as always yielding an unknown result.
In a next step, all unknown conditions can be removed.
This introduces an over- and underapproximation ruleset, called upper/lower closure. 
Guarantees about the original ruleset dropping/allowing a packet can be given by using the respective closure ruleset.

To summarize, we provide the following contributions: 
\begin{enumerate}
  \item a formal semantics of \iptables{} packet filtering (Sect.~\ref{sec:iptables-semantics}),
  \item chain unfolding: transforming a ruleset in the complex chain model to a ruleset in the simple list model (Sect.~\ref{sec:unfolding}),
  \item an embedded semantics with ternary logic, supporting arbitrary match conditions, introducing a lower/upper closure of accepted packets (Sect.~\ref{sec:ternary}), and
  \item normalization and translation of complex logical expressions to an \iptables{}-compatible format, discovering a meta-logical firewall algebra (Sect.~\ref{sec:normalization}).
\end{enumerate}

We evaluate applicability on large real-world firewalls in Sect.~\ref{sec:evaluation}.
All proofs are machine-verified with Isabelle~\cite{isabelle2014} (Sect.~\ref{sec:isabelle}).
Therefore, the correctness of all obtained results only depends on a small and well-established mathematical kernel and the \iptables{} semantics (Fig.~\ref{fig:semantics}).

\section{Firewall Models in the Literature and Related Work}
\label{sec:firewall-models}

Packets are routed through the firewall and the firewall needs to decide whether to allow or deny a packet.
A firewall ruleset determines the firewall's filtering behavior.
The firewall inspects its ruleset for each single, arbitrary packet to determine the action to apply to the packet.
The ruleset can be viewed as a list of rules; usually it is processed sequentially and the first matching rule is applied.

The literature agrees on the definition of a single firewall rule.
It consists of a predicate (the match expression) and an action.
If the match expression applies to a packet, the action is performed.
%
%
%
%
Usually, a packet is scrutinized by several rules. 
Zhang~\etal\cite{ZhangAlShaer2007flip} specify a common format for packet filtering rules.
The action is either ``allow'' or ``deny'', which directly corresponds to the firewall's filtering decision.
The ruleset is processed strictly sequentially.
Yuan \etal\cite{fireman2006} call this the \emph{simple list model}.
ITVal also supports calls to user-defined chains as an action.
This allows ``jumping'' within the ruleset without having a final filtering decision yet.
This is called the \emph{complex chain model}~\cite{fireman2006}.

In general, a packet header is a bitstring which can be matched against~\cite{zhang2012qbfsatfirewalls}.
Zhang~\etal\cite{ZhangAlShaer2007flip} support matching on the following packet header fields: 
IP source and destination address, protocol, and port on layer 4.
This model is commonly found in the literature~\cite{cspfirewall,bartal1999firmato,ZhangAlShaer2007flip,fireman2006,brucker2008modelfwisabelle}.
ITVal extends these match conditions with flags (\eg \texttt{TCP SYN}) and connection states (\texttt{INVALID}, \texttt{NEW}, \texttt{ESTABLISHED}, \texttt{RELATED)}.
The state matching is treated as just another match condition.%
\endnote{Firewalls can be stateful or stateless. 
Most firewalls nowadays are stateful, which means the firewall remembers and tracks information of previously seen packets, \eg the TCP connection a packet belongs to and the state of this connection. 
ITVal does not track the state of connections. Match conditions on connection states are treated exactly the same as matches on a packet header. 
In general, focusing on rulesets and not firewall implementation, matching on \iptables{} conntrack states is exactly as matching on any other (stateless) condition. 
However, internally, not only the packet header is consulted but also the current connection tables. 
Note that existing firewall analysis tools also largely ignore state~\cite{nelson2010margrave}.
In our semantics, we also model stateless matching. } 
This model is similar to Margrave's model for IOS~\cite{nelson2010margrave}.
When comparing these features to the simple firewall in Fig.~\ref{fig:firewall:synology}, it becomes obvious that none of these tools supports that firewall.

We are not aware of any tool which uses a model fundamentally different than those described in the previous paragraph. 
Our model enhances existing work in that we use ternary logic to support arbitrary match conditions.
To analyze a large \iptables{} firewall, the authors of Margrave~\cite{nelson2010margrave} translated it to basic Cisco IOS access lists\cite{ciscofirewallaccesslists} by hand.
With our simplification, we can automatically remove all features not understood by basic Cisco IOS.
This enables translation of any \iptables{} firewall to a basic Cisco access lists which is guaranteed to drop no more packets than the original \iptables{} firewall. 
This opens up all tools available only for Cisco IOS access lists, \eg Margrave \cite{nelson2010margrave} and Header Space Analysis~\cite{kazemian2012HSA}.%
\endnote{Note that the other direction is considered easy~\cite{ciscotoiptables}, because basic Cisco IOS access lists have ``no nice features''~\cite{ciscononicefeatures}. Note that there also are \emph{Advanced} Access Lists. }

\section{Formal Verification with Isabelle}
\label{sec:isabelle}
We verified all proofs with Isabelle, using its standard Higher-Order Logic (HOL).
The corresponding theory files are publicly available. 
An interested reader may consult the detailed (100+ pages) proof document.

\textit{Notation. }
We use pseudo code close to SML and Isabelle.
Function application is written without parentheses, \eg $f\ a$ denotes function $f$ applied to parameter $a$.
We write $\lstcons$ for prepending a single element to a list, \eg $a \lstcons b \lstcons [c,\,d] = [a,\,b,\,c,\,d]$, and $\lstapp$ for appending lists, \eg $[a,\,b] \lstapp [c,\,d] = [a,\,b,\,c,\,d]$.
The empty list is written as $[]$.
$[f\ a.\ a \leftarrow l]$ denotes a list comprehension, \ie applying $f$ to every element $a$ of list $l$.
$[f\ x\ y.\ \ x \leftarrow l_1,\; y \leftarrow l_2]$ denotes the list comprehension where $f$ is applied to each combination of elements of the lists $l_1$ and $l_2$.
For $f\ x \ y = (x,\,y)$, this returns the cartesian product of $l_1$ and $l_2$.

\section{Semantics of \iptables{}}
\label{sec:iptables-semantics}
We formalized the semantics of a subset of \iptables{}. 
The semantics focuses on access control, which is done in the \texttt{INPUT}, \texttt{OUTUT}, and \texttt{FORWARD} chain.
Thus packet modification (\eg NAT) is not considered (and also not allowed in these chains). 

Match conditions, \eg \texttt{source 192.168.0.0/24} and \texttt{protocol TCP}, are called \emph{primitives}.
A primitive matcher $\gamma$ decides whether a packet matches a primitive.
Formally, based on a set $X$ of primitives and a set of packets $P$, a primitive matcher $\gamma$ is a binary relation over $X$ and $P$.
The semantics supports arbitrary packet models and match conditions, hence both remain abstract in our definition.

In one firewall rule, several primitives can be specified.
Their logical connective is conjunction, for example $\verb~src 192.168.0.0/24~ \ \mathit{and} \  \verb~tcp~$.
Disjunction is omitted because it is neither needed for the formalization nor supported by \iptables{}; this is consistent with the model by Jeffrey and Samak~\cite{fwmodelchecking2009jeffry}.
Primitives can be combined in an algebra of \emph{match expressions} $M_X$:
\begin{IEEEeqnarray*}{c}
    \mathit{mexpr} \quad = \quad  x \quad \text{for } x \in X 
                   \quad | \quad  \neg\, \mathit{mexpr} 
                   \quad | \quad \mathit{mexpr} \,\wedge\, \mathit{mexpr} 
                   \quad | \quad \mathtt{True}
\end{IEEEeqnarray*}
For a primitive matcher $\gamma$ and a match expression $m \in M_X$, we write \mbox{$m \matchop\gamma p$} if a packet $p \in P$ matches $m$, essentially lifting $\gamma$ to a relation over $M_X$ and $P$, with the connectives defined as usual.
With completely generic $P$, $X$, and $\gamma$, the semantics can be considered to have access to an oracle which understands all possible match conditions.

Furthermore, we support the following \emph{actions}, modeled closely after \iptables{}: $\mathtt{Accept}$, $\mathtt{Reject}$, $\mathtt{Drop}$, $\mathtt{Log}$, $\mathtt{Empty}$, $\mathtt{Call}\ c$ $\text{for a chain $c$}$, and $\mathtt{Return}$.
A \emph{rule} can be defined as a tuple $(m,\,a)$ for a match expression $m$ and an action $a$.
A list (or sequence) of rules is called a \emph{chain}.
For example, the beginning of the \verb~DOS_PROTECT~ chain in Fig.~\ref{fig:firewall:synology} is $[(\verb~icmp~ \wedge \verb~icmptype 8 limit:~\,\dots,\: \mathtt{Return}),\: \dots]$
.

A set of chains associated with a name is called a \emph{ruleset}.
Let $\Gamma$ denote the mapping from chain names to chains.
For example, $\Gamma \ \verb~DOS_PROTECT~$ returns the contents of the \verb~DOS_PROTECT~ chain.
We assume that $\Gamma$ is well-formed that means, if a $\mathtt{Call}\ c$ action occurs in a ruleset, then the chain named $c$ is defined in $\Gamma$.
This assumption is justified as the Linux kernel only accepts well-formed rulesets.


\begin{figure}[htb]
\begin{adjustwidth}{-2em}{-2em}%
\centering
  \begin{mathpar}
    \textsc{Skip}\quad\inferrule{ }{\bigstep{[]}{t}{t}} \and
    \textsc{Accept}\quad\inferrule{\matches{m}}{\bigstep{[(m,\:\mathtt{Accept})]}{\undecided}{\allow}} \and
    \textsc{Drop}\quad\inferrule{\matches{m}}{\bigstep{[(m,\:\mathtt{Drop})]}{\undecided}{\deny}} \and
    \textsc{Reject}\quad\inferrule{\matches{m}}{\bigstep{[(m,\:\mathtt{Reject})]}{\undecided}{\deny}} \and
    \textsc{NoMatch}\quad\inferrule{\nmatches{m}}{\bigstep{[(m,\:a)]}{\undecided}{\undecided}} \and
    \textsc{Decision}\quad\inferrule{t \neq \undecided}{\bigstep{\mathit{rs}}{t}{t}} \and
    \textsc{Seq}\quad\inferrule{\bigstep{\mathit{rs}_1}{\undecided}{t} \\ \bigstep{\mathit{rs}_2}{t}{t'}}{\bigstep{\mathit{rs}_1 \lstapp \mathit{rs}_2}{\undecided}{t'}} \and
    \textsc{CallResult}\quad\inferrule{\matches{m} \\ \\ \bigstep{\Gamma\ c}{\undecided}{t}}{\bigstep{[\left(m,\: \mathtt{Call}\ c\right)]}{\undecided}{t}}\and
    \textsc{CallReturn}\quad\inferrule{\matches{m} \\ \Gamma\; c = \mathit{rs}_1 \lstapp (m',\: \mathtt{Return}) \lstcons \mathit{rs}_2 \\ \matches{m'} \\ \bigstep{\mathit{rs}_1}{\undecided}{\undecided}}{\bigstep{[\left(m,\: \mathtt{Call}\ c\right)]}{\undecided}{\undecided}} \and
    \textsc{Log}\quad\inferrule{\matches{m}}{\bigstep{[(m,\:\mathtt{Log})]}{\undecided}{\undecided}} \and
    \textsc{Empty}\quad\inferrule{\matches{m}}{\bigstep{[(m,\:\mathtt{Empty})]}{\undecided}{\undecided}}
  \end{mathpar}
\end{adjustwidth}
  \hfill (for any primitive matcher $\gamma$ and any well-formed ruleset $\Gamma$)
  \caption{Big Step semantics for \iptables{}}
  \label{fig:semantics}
\end{figure}

The semantics of a firewall w.r.t.\ to a given packet $p$, a background ruleset $\Gamma$, and a primitive matcher $\gamma$ can be defined as a relation over the currently active chain and the state before and the state after processing this chain.
The semantics is specified in Fig.~\ref{fig:semantics}.
The expression $\bigstep{\mathit{rs}}{t}{t'}$ states that starting with state $t$, after processing the chain $\mathit{rs}$, the resulting state is $t'$.
For a packet $p$, our semantics focuses on firewall filtering decisions.
Therefore, only the following three states are necessary:	
The firewall may allow ($\allow$) or deny ($\deny$) the packet, or it may not have come to a decision yet~($\undecided$). 

We will now discuss the most important rules. 
The \textsc{Accept} rule describes the following: if the packet $p$ matches the match expression $m$, then the firewall with no filtering decision ($\undecided$) processes the singleton chain $[(m,\:\mathtt{Accept})]$ by switching to the allow state.
Both the \textsc{Drop} and \textsc{Reject} rules deny a packet; the difference is only in whether the firewall generates some informational message, which does not influence filtering. 
The \textsc{NoMatch} rule specifies that if the firewall has not come to a filtering decision yet, it can process any non-matching rule without changing its state.
The \textsc{Decision} rule specifies that as soon as the firewall made a filtering decision, it does not change its decision.
The \textsc{Seq} rule specifies that if the firewall has not come to a filtering decision and it processes the chain $\mathit{rs}_1$ which results in state $t$ and starting from $t$ processes the chain $\mathit{rs}_2$ which results in state $t'$, then both chains can be processed sequentially, ending in state $t'$. 
The \textsc{CallResult} rule specifies that if a matching $\mathtt{Call}$ to a chain named ``$c$'' occurs, the resulting state $t$ is the result of processing the chain $\Gamma\; c$.
Likewise, the \textsc{CallReturn} rule specifies that if processing a prefix $\mathit{rs}_1$ of the called chain does not lead to a filtering decision and directly afterwards, a matching $\mathtt{Return}$ rule occurs, the called chain is processed without result.%
\endnote{The semantics gets stuck if a $\mathtt{Return}$ occurs on top-level.
However, this is not a problem since we make sure that this cannot happen.
\iptables{} specifies that a $\mathtt{Return}$ on top-level in a built-in chain is allowed and in this corner case, the chain's default policy is executed. 
To comply with this behavior, we always start analysis of a ruleset as follows: 
$[(\mathtt{True},\: \mathtt{Call}\ \mathit{start\mhyphen{}chain}),\allowbreak{}\: (\mathtt{True},\: \mathit{default\mhyphen{}policy})]$, where the start chain is one of \iptables{}' built-in \texttt{INPUT}, \texttt{FORWARD}, or \texttt{OUTPUT} chains with a default policy of either \texttt{Accept} or \texttt{Drop}. }
The \textsc{Log} rule does not influence the filtering behavior.
Similarly, the \textsc{Empty} rule does not result in a filtering decision.
An \textsc{Empty} rule, \ie a rule without an action, occurs if \iptables{} only updates its internal state, \eg updating packet counters.\endnote{%
A rule without an action can also be used to mark a packet for later handling.
This marking may influence the filtering decision.
Since our primitive matchers and packets are completely generic, this case can be represented within our model:
Instead of updating the firewall's internal state, an additional ``ghost field'' must be introduced in the packet model.
Since packets are immutable, this field cannot be set by a rule but the packet must be given to the firewall with the final value of the ghost field already set.
Hence, an analysis must be carried out with an arbitrary value of the ghost fields.
We admit that this model is very unwieldy.
However, when later embedding the more practical ternary semantics, we want to mention that all primitives which mark a packet for later processing can be considered ``unknown'' and are correctly abstracted by these semantics. 
}

The subsequent parts of this paper are all based on these semantics.
Whenever we provide a procedure $P$ to operate on chains, we proved that the firewall's filtering behavior is preserved, formally:%
\begin{IEEEeqnarray*}{c}
\bigstep{P \ \mathit{rs}}{t}{t'} \quad \text{\textit{iff}} \quad \bigstep{\mathit{rs}}{t}{t'}
\end{IEEEeqnarray*}%
All our proofs are machine-verified with Isabelle.
Therefore, once the reader is convinced of the semantics as specified in Fig.~\ref{fig:semantics}, the correctness of all subsequent theorems follows automatically -- without any hidden assumptions or limitations. 

The rules in Fig.~\ref{fig:semantics} are designed such that every rule can be inspected individually.
However, considering all of them together, it is not immediately clear whether the result depends on the order of their application to a concrete ruleset and packet.
Theorem~\ref{thm:bigstep-deterministic} states that the semantics is deterministic, \ie only one uniquely defined outcome is possible.

\begin{theorem}[Determinism]%
\label{thm:bigstep-deterministic}%
\thmskip{}%
\begin{IEEEeqnarray*}{c}
\text{If} \quad \bigstep{\mathit{rs}}{s}{t} \quad \text{and} \quad \bigstep{\mathit{rs}}{s}{t'} \quad \text{then} \quad t=t'
\end{IEEEeqnarray*}
\end{theorem}

\section{Custom Chain Unfolding}
\label{sec:unfolding}
In this section, we present algorithms to convert a ruleset from the complex chain model to the simple list model.

The function \texttt{pr} (``process return'') iterates over a chain.
If a \texttt{Return} rule is encountered, all subsequent rules are amended by adding the \texttt{Return} rule's negated match expression as a conjunct.
Intuitively, if a \texttt{Return} rule occurs in a chain, all following rules of this chain can only be reached if the \texttt{Return} rule does not match.
\begin{IEEEeqnarray*}{lCl}
  \mathtt{add\mhyphen{}match} \ m' \ \mathit{rs} & = & [(m \wedge m',\,a).\ (m,\,a) \leftarrow \mathit{rs}] \\
  \mathtt{pr} \ [] & = & [] \\
  \mathtt{pr} \ ((m,\, \mathtt{Return}) \lstcons \mathit{rs}) & = & \mathtt{add\mhyphen{}match} \ (\neg m) \ (\mathtt{pr} \ \mathit{rs})\\
  \mathtt{pr} \ ((m,\, a) \lstcons \mathit{rs}) & = & (m,\, a) \lstcons \mathtt{pr} \ \mathit{rs}
\end{IEEEeqnarray*}

The function \texttt{pc} (``process call'') iterates over a chain, unfolding one level of $\texttt{Call}$ rules.
If a \texttt{Call} to the chain $c$ occurs, the chain itself (\ie $\Gamma \ c$) is inserted instead of the \texttt{Call}.
However, \texttt{Return}s in the chain need to be processed and the match expression for the original \texttt{Call} needs to be added to the inserted chain.
\begin{IEEEeqnarray*}{lCl}
  \mathtt{pc} \ [] & = & []\\
  \mathtt{pc} \ ((m,\, \mathtt{Call}\ c) \lstcons \mathit{rs}) & = & \mathtt{add\mhyphen{}match} \  m \ \left(\mathtt{pr} \ \left(\Gamma \ c\right)\right) \lstapp \mathtt{pc} \ \mathit{rs} \\
  \mathtt{pc} \ ((m,\, a) \lstcons \mathit{rs}) & \ = \ & (m,\, a) \lstcons \mathtt{pc} \ \mathit{rs}
\end{IEEEeqnarray*}

The procedure \texttt{pc} can be applied arbitrarily many times and preserves the semantics.
It is sound and complete.
\begin{theorem}[Soundness and Completeness]%
\label{thm:unfolding-sound-complete}%
\thmskip{}%
\begin{IEEEeqnarray*}{c}
   \bigstep{\mathtt{pc}^n \ \mathit{rs}}{t}{t'} \quad  \text{iff} \quad  \bigstep{\mathit{rs}}{t}{t'}
\end{IEEEeqnarray*}
\end{theorem}

In each iteration, the algorithm unfolds one level of $\mathtt{Call}$s.
The algorithm needs to be applied until the result no longer changes.
Note that the semantics allows non-terminating rulesets; however, the only rulesets that are interesting for analysis are the ones actually accepted by the Linux kernel.\endnote{The relevant check is in \texttt{mark\_source\_chains}, file \url{source/net/ipv4/netfilter/ip_tables.c} of the Linux kernel version 3.2.} 
%
Since it rejects rulesets with loops, both our algorithm and the resulting ruleset are guaranteed to terminate.

\begin{corollary}
Every ruleset (with only $\mathtt{Accept}$, $\mathtt{Drop}$, $\mathtt{Reject}$, $\mathtt{Log}$, $\mathtt{Empty}$, $\mathtt{Call}$, $\mathtt{Return}$ actions) accepted by the Linux kernel can be unfolded completely while preserving its filtering behavior.
\end{corollary}

In addition to unfolding calls, the following transformations applied to any ruleset preserve the semantics:
\begin{itemize}
  \item 
    Replacing $\mathtt{Reject}$ actions with $\mathtt{Drop}$ actions, 
  \item 
    Removing $\mathtt{Empty}$ and $\mathtt{Log}$ rules, 
  \item 
    Simplifying match expressions which contain \texttt{True} or $\neg\,\mathtt{True}$.
   \item 
   For some given primitive matcher, specific optimizations may also be performed, \eg rewriting \texttt{src 0.0.0.0/0} to \texttt{True}.
\end{itemize}

Therefore, after unfolding and optimizing, a chain which only contains $\mathtt{Allow}$ or $\mathtt{Drop}$ actions is left.
In the subsequent sections, we require this as a precondition.
As an example, recall the firewall in Fig.~\ref{fig:firewall:synology}.
Its \verb~INPUT~ chain after unfolding and optimizing is listed in Fig. ~\ref{fig:firewall:synology-unfolded}.
%
%
Observe that the computed match expressions are beyond iptable's expressiveness.
An algorithm to normalize the rules to an \iptables{}-compatible format will be described in Sect.~\ref{sec:normalization}.

\begin{figure}[t]%
\begin{IEEEeqnarray*}{l}%
[           \left(\neg\,\left(\verb~icmp~ \wedge \verb~icmptype 8 limit:~\dots\right) \,\wedge \verb~icmp~ \wedge \verb~icmptype 8~,\, \mathtt{Drop}\right),\;\\
\phantom{[} (\neg\,\left(\verb~icmp~ \wedge \verb~icmptype 8 limit:~\dots\right) \,\wedge \neg\,\left(\verb~tcp~ \wedge \verb~tcp flags:0x17/0x04 limit:~ \dots\right) \,\wedge \\
\phantom{[( } \verb~tcp~ \wedge \verb~tcp flags:0x17/0x04~,\,  \mathtt{Drop}),\; \dots, \;
\phantom{[} \left(\verb~src 192.168.0.0/16~,\, \mathtt{Accept}\right),\; \dots ]%
\end{IEEEeqnarray*}%
\vskip-10pt 
\caption{Unfolded Synology Firewall}%
  \label{fig:firewall:synology-unfolded}%
\end{figure}

\section{Unknown Primitives}
\label{sec:ternary}
As we argued earlier, it is infeasible to support all possible primitives of a firewall.
Suppose a new firewall module is created which provides the \verb~ssh_blacklisted~ and \verb~ssh_innocent~ primitives. 
The former applies if an IP address has had too many invalid SSH login attempts in the past; 
the latter is the opposite of the former.
Since we made up these primitives, no existing tool will support them. 
However, a new version of \iptables{} could implement them or they can be provided as third-party kernel modules. 
Therefore, our ruleset transformations must take unknown primitives into account. 
To achieve this, we lift the primitive matcher $\gamma$  to ternary logic, adding $\mathtt{Unknown}$ as matching outcome. 
We embed this new ``approximate'' semantics into the semantics described in the previous sections.
Thus, it becomes easier to construct matchers tailored to the primitives supported by a particular tool.

\subsection{Ternary Matching}

Logical conjunction and negation on ternary values are as before, with these additional rules for \texttt{Unknown} operands (commutative cases omitted):
\begin{IEEEeqnarray*}{c}
    \mathtt{True}  \wedge  \mathtt{Unknown} = \mathtt{Unknown} \ \quad
    \mathtt{False}  \wedge  \mathtt{Unknown} = \mathtt{False} \ \quad
                      \neg \: \mathtt{Unknown} = \mathtt{Unknown}
\end{IEEEeqnarray*}
These rules correspond to Kleene's 3-valued logic~\cite{kleene1952introduction} and are well-suited for firewall semantics: 
The first equation states that, if one condition matches, the final result only depends on the other condition.
The next equation states that a rule cannot match if one of its conditions does not match.
Finally, by negating an unknown value, no additional information can be inferred.

We demonstrate this by example: 
the two rulesets
  $\left[(\texttt{ssh\_blacklisted},\, \mathtt{Drop})\right]$
and
$ \left[(\mathtt{True},\, \mathtt{Call}\ c)\right]$
  where $\Gamma\, c\, = [(\mathtt{ssh\_innocent},\, \mathtt{Return}),\, (\mathtt{True},\, \mathtt{Drop})]$
have exactly the same filtering behavior.
After unfolding, the second ruleset collapses to
 $ \left[(\neg\; \texttt{ssh\_innocent},\, \texttt{Drop})\right]$.
Both the \texttt{ssh\_blacklisted} and the \texttt{ssh\_innocent} primitives are \texttt{Unknown} to our matcher.
Thus, since both rulesets have the same filtering behavior, a packet matching $\mathtt{Unknown}$ in the first ruleset should also match $\neg\;\mathtt{Unknown}$ in the second ruleset matches.


%

\subsection{Closures}
In the ternary semantics, it may be unknown whether a rule applies to a packet.
Therefore, the matching semantics are extended with an \emph{``in-doubt''-tactic}.
This tactic is consulted if the result of a match expression is \texttt{Unknown}.
It decides whether a rule applies.

We introduce the \emph{in-doubt-$\mathit{allow}$} and \emph{in-doubt-$\mathit{deny}$} tactics.
The first tactic forces a match if the rule's action is \texttt{Accept} and a mismatch if it is \texttt{Drop}.
The second tactic behaves in the opposite manner.
Note that an unfolded ruleset is necessary, since no behavior can be specified for \texttt{Call} and \texttt{Return} actions.\endnote{The final decision ($\allow$ or $\deny$) for \texttt{Call} and \texttt{Return} rules depends on the called/calling chain.}

We denote the exact Boolean semantics with ``$\Rightarrow$'' and embedded ternary semantics with an arbitrary tactic $\alpha$ with ``$\Rightarrow_\alpha$''.
In particular, $\alpha = \mathit{allow}$ for \emph{in-doubt-$\mathit{allow}$} and $\alpha = \mathit{deny}$ analogously.

``$\Rightarrow$'' and ``$\Rightarrow_\alpha$'' are related to the in-doubt-tactics as follows: 
considering the set of all accepted packets, \emph{in-doubt-$\mathit{allow}$} is an overapproximation, whereas \emph{in-doubt-$\mathit{deny}$} is an underapproximation.
In other words, if ``$\Rightarrow$'' accepts a packet, then ``$\Rightarrow_{\operatorname{allow}}$'' also accepts the packet.
Thus, from the opposite perspective, the \emph{in-doubt-$\mathit{allow}$} tactic can be used to guarantee that a packet is certainly dropped.
Likewise, if ``$\Rightarrow$'' denies a packet, then ``$\Rightarrow_{\operatorname{deny}}$'' also denies this packet.
Thus, the \emph{in-doubt-$\mathit{deny}$} tactic can be used to guarantee that a packet is certainly accepted.

For example, the unfolded firewall of Fig.~\ref{fig:firewall:synology} contains rules which drop a packet if a limit is exceeded.
If this rate limiting is not understood by $\gamma$, the \emph{in-doubt-$\mathit{allow}$} tactic will never apply this rule, while with the \emph{in-doubt-$\mathit{deny}$} tactic, it is applied universally.

We say that the Boolean and the ternary matchers agree iff they return the same result or the ternary matcher returns \texttt{Unknown}.
Interpreting this definition, the ternary matcher may always return \texttt{Unknown} and the Boolean matcher serves as an oracle which knows the correct result.
Note that we never explicitly specify anything about the Boolean matcher; therefore the model is universally valid, \ie the proof holds for an arbitrary oracle.

If the exact and ternary matcher agree, then the set of all packets allowed by the \emph{in-doubt-$\mathit{deny}$} tactic is a subset of the packets allowed by the exact semantics, which in turn is a subset of the packets allowed by the \emph{in-doubt-$\mathit{allow}$} tactic.
Therefore, we call all packets accepted by $\Rightarrow_{\operatorname{deny}}$ the \emph{lower closure}, \ie the semantics which accepts at most the packets that the exact semantics accepts.
Likewise, we call all packets accepted by $\Rightarrow_{\operatorname{allow}}$ the \emph{upper closure}, \ie the semantics which accepts at least the packets that the exact semantics accepts.
Every packet which is not in the upper closure is guaranteed to be dropped by the firewall.

\begin{theorem}[Lower and Upper Closure of Allowed Packets]%
\label{thm:FinalAllowClosure}%
\thmskip{}%
\begin{IEEEeqnarray*}{c}
  \left\lbrace p.\ \bigstepapprox{\mathit{rs}}{\undecided}{\allow}{\operatorname{deny}} \right\rbrace 
  \subseteq\,  \left\lbrace p.\ \bigstep{\mathit{rs}}{\undecided}{\allow} \right\rbrace 
  \subseteq\,  \left\lbrace p.\ \bigstepapprox{\mathit{rs}}{\undecided}{\allow}{\operatorname{allow}} \right\rbrace
\end{IEEEeqnarray*}
\end{theorem}

The opposite holds for the set of denied packets.

For the example in Fig.~\ref{fig:firewall:synology}, we computed the closures (without the \texttt{RELATED\hskip-1.2pt{},\allowbreak{}ESTABLISHED} rule, see Sect.~\ref{sec:establishedrule}) and a ternary matcher which only understands IP addresses and layer 4 protocols.
The lower closure is the empty set since rate limiting could apply to any packet.
The upper closure is the set of packets originating from $192.168.0.0/16$.

\subsection{Removing Unknown Matches}
In this section, as a final optimization, we remove all unknown primitives.
We call this algorithm \texttt{pu} (``process unknowns'').
For this step, the specific ternary matcher and the choice for the in-doubt-tactic must be known.

In every rule, top-level unknown primitives can be rewritten to $\texttt{True}$ or $\neg\, \mathtt{True}$.
For example, let $m_u$ be a primitive which is unknown to $\gamma$. Then, for in-doubt-allow, $(m_u,\, \mathtt{Accept})$ is equal to $(\mathtt{True},\, \mathtt{Accept})$ and $(m_u,\, \mathtt{Drop})$ is equal to $(\neg\,\mathtt{True},\, \mathtt{Drop})$.
Similarly, negated unknown primitives and conjunctions of (negated) unknown primitives can be rewritten.

Hence, the base cases of \texttt{pu} are straightforward.
However, the case of a negated conjunction of match expressions requires some care.
The following equation represents the De Morgan rule, specialized to the in-doubt-allow tactic.
\begin{IEEEeqnarray*}{l}
\textnormal{\texttt{pu}} \ (\neg\,(m_1 \wedge m_2),\; a) \quad = \quad
  \begin{cases}
    \mathtt{True} & \qquad\hfill \text{if } \mathtt{pu} \ (\neg\,m_1,\; a) = \mathtt{True}  \phantom{\neg\,} \\
    \mathtt{True} & \qquad\hfill \text{if } \mathtt{pu}  \ (\neg\,m_2,\; a) = \mathtt{True} \phantom{\neg\,} \\
    \mathtt{pu} \ (\neg\,m_2,\; a) & \qquad\hfill \text{if } \mathtt{pu}  \ (\neg\,m_1,\; a) = \neg\,\mathtt{True} \\
    \mathtt{pu} \ (\neg\,m_1,\; a) & \qquad\hfill \text{if } \mathtt{pu}  \ (\neg\,m_2,\; a) = \neg\,\mathtt{True} \\
    \multicolumn{2}{l}{ \neg\left(\neg\,\mathtt{pu}\; \left(\neg\,m_1,\, a\right) \wedge \neg\,\mathtt{pu}\; \left(\neg\,m_2,\, a\right)\right)  \qquad \text{otherwise} }
  \end{cases}
\end{IEEEeqnarray*}


The $\neg\;\mathtt{Unknown} = \mathtt{Unknown}$ equation is responsible for the complicated nature of the De Morgan rule.
Fortunately, we machine-verified all our algorithms. 
For example, during our research, we wrote a seemingly simple (but incorrect) version of \texttt{pu} and everybody agreed that the algorithm looks correct.
In the early empirical evaluation, with yet unfinished proofs, we did not observe our bug.
Only because of the failed correctness proof did we realize that we introduced an equation that only holds in Boolean logic.

\begin{theorem}[Soundness and Completeness]%
\thmskip{}%
\begin{IEEEeqnarray*}{c}
\bigstepapprox{[\textnormal{\texttt{pu}} \ r.\ \ r \leftarrow \ \mathit{rs}]}{t}{t'}{\operatorname{allow}} \quad \text{iff} \quad \bigstepapprox{\mathit{rs}}{t}{t'}{\operatorname{allow}}
\end{IEEEeqnarray*}
\end{theorem}
\begin{theorem}%
Algorithm \textnormal{\texttt{pu}} removes all unknown primitive match expressions. 
\end{theorem}

An algorithm for the in-doubt-deny tactic (with the same equation for the De Morgan case) can be specified in a similar way.
Thus, $\Rightarrow_{\alpha}$ can be treated as if it were defined only on Boolean logic with only known match expressions.

As an example, we examine the ruleset of the upper closure of Fig.~\ref{fig:firewall:synology} (without the \texttt{RELATED\hskip-1.2pt{},\allowbreak{}ESTABLISHED} rule, see Sect.~\ref{sec:establishedrule}) for a ternary matcher which only understands IP addresses and layer 4 protocols.
The ruleset is simplified to $[(\verb~src 192.168.0.0/16~,\, \mathtt{Accept}),\allowbreak{}\, (\mathtt{True},\, \mathtt{Drop})]$. 
ITVal can now directly compute the correct results on this ruleset.

\subsection{The \texttt{RELATED\hskip-1.2pt{},ESTABLISHED} Rule}
\label{sec:establishedrule}
Since firewalls process rules sequentially, the first rule has no dependency on any previous rules.
Similarly, rules at the beginning have very low dependencies on other rules.
Therefore, firewall rules in the beginning can be inspected manually, whereas the complexity of manual inspection increases with every additional preceding rule.

It is good practice~\cite{iptablesperfectruleset} to start a firewall with an \texttt{ESTABLISHED} (and sometimes \texttt{RELATED}) rule.
This also happens in Fig.~\ref{fig:firewall:synology} after the rate limiting.
The \texttt{ESTABLISHED} rule usually matches most of the packets~\cite{iptablesperfectruleset},\endnote{We revalidated this observation in September 2014 and found that in our firewall, which has seen more than 15 billion packets (19+ Terabyte data) since the last reboot, more than 95\% of all packets matched the first \texttt{RELATED\hskip-1.2pt{},\allowbreak{}ESTABLISHED} rule. } which is important for performance; however, when analyzing the filtering behavior of a firewall, it is important to consider how a connection can be brought to this state.
Therefore, we remove this rule and only focus on the connection setup. 

The \texttt{ESTABLISHED} rule essentially allows packet flows in the opposite direction of all subsequent rules \cite{diekmann2014EPTCS}. 
Unless there are special security requirements (which is not the case in any of our analyzed scenarios), the \texttt{ESTABLISHED} rule can be excluded when analyzing the connection setup~\cite[Corollary 1]{diekmann2014EPTCS}.\endnote{%
The same can be concluded for reflexive ACLs in Cisco's IOS Firewall~\cite{ciscofirewallaccesslists}.} 
If the \texttt{ESTABLISHED} rule is removed and in the subsequent rules, for example, a primitive \texttt{state NEW} occurs, our ternary matcher returns \texttt{Unknown}.
The closure procedures handle these cases automatically, without the need for any additional knowledge.

\section{Normalization}
\label{sec:normalization}
Ruleset unfolding may result in non-atomic match expressions, \eg $\neg\:(a \wedge b)$.
\iptables{} only supports match expressions in \emph{Negation Normal Form} (NNF).\endnote{
Since match expressions do not contain disjunctions, any match expression in NNF is trivially also in \emph{Disjunctive Normal Form} (DNF). }
There, a negation may only occur before a primitive, not before compound expressions.
For example, $\neg\:(\verb~src ~\mathit{ip}) \:\wedge\: \mathtt{tcp}$ is a valid NNF formula, whereas $\neg\:\left(\left(\verb~src ~\mathit{ip}\right) \:\wedge\: \mathtt{tcp}\right)$ is not.
We normalize match expressions to NNF, using the following observations:

The De Morgan rule can be applied to match expressions, splitting one rule into two.
For example, $\left[\left(\neg\;(\verb~src ~\mathit{ip} \;\wedge\; \verb~tcp~),\, \mathtt{Allow}\right)\right]$ and $[(\neg\;\verb~src ~\mathit{ip},\; \mathtt{Allow}),\allowbreak \, (\neg\; \verb~tcp~,\, \mathtt{Allow})]$ are equivalent. 
%
This introduces a ``meta-logical'' disjunction consisting of a sequence of consecutive rules with a shared action.
For example, $[(m_1,\, a),\; (m_2,\, a)]$ is equivalent to $[(m_1 \vee m_2,\, a)]$. 

For sequences of rules with the same action, a distributive law akin to common Boolean logic holds.
For example, the conjunction of the two rulesets $[(m_1,\, a),\allowbreak{}\; (m_2,\, a)]$ and $[(m_3,\, a),\allowbreak{}\; (m_4,\, a)]$ is equivalent to the ruleset $[({m_1 \wedge m_3},\allowbreak{}\, a),\allowbreak{}\; ({m_1 \wedge m_4},\, a),\; (m_2 \wedge m_3,\, a),\; (m_2 \wedge m_4,\, a)]$.
This can be illustrated with a situation where $a = \mathtt{Accept}$ and a packet needs to pass two firewalls in a row.

We can now construct a procedure which converts a rule with a complex match expression to a sequence of rules with match expressions in NNF. 
It is independent of the particular primitive matcher and the in-doubt tactic used. 
The algorithm \texttt{n} (``normalize'') of type $M_X \to \operatorname{List}(M_X)$ is defined as follows:
\begin{IEEEeqnarray*}{lcl}
  \textnormal{\texttt{n}} \ \mathtt{True} & \ = \ & [\mathtt{True}]\\
  \textnormal{\texttt{n}} \ (m_1 \wedge m_2) & \ = \ & [x \wedge y.\ \ x \leftarrow \textnormal{\texttt{n}}\ m_1,\; y \leftarrow \textnormal{\texttt{n}}\ m_2]\\
  \textnormal{\texttt{n}} \ (\neg\;(m_1 \wedge m_2)) & \ = \ & \textnormal{\texttt{n}}\ (\neg m_1) \ \lstapp \ \textnormal{\texttt{n}}\ (\neg m_2)\\
  \textnormal{\texttt{n}} \ (\neg\neg m) & \ = \ & \textnormal{\texttt{n}}\ m\\
  \textnormal{\texttt{n}} \ (\neg \mathtt{True}) & \ = \ & []\\
  \textnormal{\texttt{n}} \ x & \ = \ & [x] \qquad \raisebox{-2.5mm}[0pt][0pt]{\Bigg\} for $x \in X$} \\
  \textnormal{\texttt{n}} \ (\neg x) & \ = \ & [\neg x]
\end{IEEEeqnarray*}
The second equation corresponds to the distributive law, the third to the De Morgan rule.
For example, $\texttt{n}\ \left(\neg\,(\verb~src~\;\mathit{ip} \wedge \verb~tcp~)\right) = \left[\neg\,\verb~src~\;\mathit{ip},\, \neg\, \verb~tcp~\right]$.
The fifth rule states that non-matching rules can be removed completely.

The unfolded ruleset of Fig.~\ref{fig:firewall:synology-unfolded}, which consists of $9$ rules, can be normalized to a ruleset of $20$ rules (due to distributivity).
In the worst case, normalization can cause an exponential blowup.
Our evaluation shows that this is not a problem in practice, even for large rulesets. 
This is because rulesets are usually managed manually, which naturally limits their complexity to a level processible by state-of-the-art hardware. 

\begin{theorem}
  \textnormal{\texttt{n}} always terminates, all match expressions in the returned list are in NNF, and their conjunction is equivalent to the original expression.
\end{theorem}

We show soundness and completeness w.r.t.\ arbitrary $\gamma$, $\alpha$, and primitives.
Hence, it also holds for the Boolean semantics.
In general, proofs about the ternary semantics are stronger (the ternary primitive matcher can simulate the Boolean matcher).

\pagebreak
\begin{theorem}[Soundness and Completeness]%
\label{thm:n-sound-complete-ternary}%
\thmskip{}%
  \begin{IEEEeqnarray*}{c}
    \bigstepapprox{[(m',\, a).\ \ m' \leftarrow\textnormal{\texttt{n}}\ m]}{t}{t'}{\alpha} 
    \quad \text{iff} \quad  \bigstepapprox{[(m,\, a)]}{t}{t'}{\alpha}
  \end{IEEEeqnarray*}
\end{theorem}


After having been normalized by \textnormal{\texttt{n}}, the rules can mostly be fed back to \iptables{}.
For some specific primitives, \iptables{} imposes additional restrictions, \eg that at most one primitive of a type may be present in a single rule. 
For our evaluation, we only need to solve this issue for IP address ranges in CIDR notation~\cite{rfc4632}. 
We introduced and verified another transformation which computes intersection of IP address ranges, which returns at most one range.
This is sufficient to process all rulesets we encountered during evaluation.

\section{Evaluation}
\label{sec:evaluation}
In this section, we demonstrate the applicability of our ruleset preprocessing. 
Usually, network administrators are not inclined towards publishing their firewall ruleset because of potential negative security implications. 
For this evaluation, we have obtained approximately $20\textnormal{k}$ real-world rules and the permission to publish them.
In addition to the running example in Fig.~\ref{fig:firewall:synology} (a small real-world firewall), we tested our algorithms on four other real-world firewalls.
We put focus on the third ruleset, because it is one of the largest and the most interesting one.

For our analysis, we wanted to know how the firewall partitions the IPv4 space.
Therefore, we used a matcher $\gamma$ which only understands source/destination IP addresses and the layer 4 protocols TCP and UDP.
Our algorithms do not require special processing capabilities, they can be executed within seconds on a common off-the-shelf \SI{4}{\giga\byte} RAM laptop.

\paragraph*{Ruleset 1 }
is taken from a Shorewall~\cite{shorewall} firewall, running on a home router, with around 500 rules.
We verified that our algorithms correctly unfold, preprocess, and simplify this ruleset.
We expected to see, in both the upper and lower closure, that the firewall drops packets from private IP ranges.
However, we could not see this in the upper closure and verified that the firewall does indeed not block such packets if their connection is in a certain state.
The administrator of the firewall confirmed this issue and is currently investigating it.

\paragraph*{Ruleset 2 }
is taken from a small firewall script found online~\cite{iptablesringofsaturn}.
Although it only contains about 50 rules,  we found that it contains a serious mistake. 
We assume the author accidentally confused \iptables{}' \texttt{-I} (insert at top) and \texttt{-A} (append at tail) options.
We saw this after unfolding, as the firewall allows nearly all packets at the beginning.
Subsequent rules are shadowed and cannot apply.
However, these rules come with a documentation of their intended purpose, such as ``drop reserved addresses'', which highlights the error.   
We verified the erroneous behavior by installing the firewall on our systems.
The author is currently investigating this issue.
Thus, our unfolding algorithm alone can provide valuable insights.

\paragraph*{Ruleset 3 \& 4 }
are taken from the main firewall of our lab (Chair for Network Architectures and Services).
One snapshot was taken 2013 with 2800 rules and one snapshot was taken 2014, containing around 4000 rules.
It is obvious that these rulesets have historically grown.
About ten years ago, these two rulesets would have been the largest real-world rulesets ever analyzed in academia~\cite{firwallerr2004}.

We present the analysis results of the 2013 version of the firewall.
Details can be found in the additional material. 
We removed the first three rules. 
The first rule was the \texttt{ESTABLISHED} rule, as discussed in Sect.~\ref{sec:establishedrule}.
Our focus was put on the second rule when we calculated the lower closure:  
this rule was responsible for the lower closure being the empty set.
Upon closer inspection of this rule, we realized that it was `dead', \ie it can never apply.
We confirmed this observation by changing the target to a $\mathtt{Log}$ action on the real firewall and could never see a hit of this rule for months.
Due to our analysis, this rule could be removed.
%
The third rule performed SSH rate limiting (a $\mathtt{Drop}$ rule).
We removed this rule because we had a very good understanding of it.
Keeping it would not influence correctness of the upper closure, but lead to a smaller lower closure than necessary.

First, we tested the ruleset with the well-maintained Firewall Builder~\cite{fwbuilder}.
The original ruleset could not be imported by Firewall Builder due to $22$ errors, caused by unknown match expressions.
Using the calculated upper closure, Firewall Builder could import this ruleset without any problems.

Next, we tested ITVal's IP space partitioning query~\cite{marmorstein2006firewall}.
On our original ruleset with 2800 rules, ITVal completed the query with around \SI{3}{\giga\byte} of RAM in around \SI{1}{\minute}.
Analyzing ITVal's debug output, we found that most of the rules were not understood correctly due to unknown primitives.
Thus, the results were spurious.
We could verify this as 127.0.0.0/8, obviously dropped by our firewall, was grouped into the same class as the rest of the Internet.
In contrast, using the upper and lower closure ruleset, ITVal correctly identifies 127.0.0.0/8 as its own class.

We found another interesting result about the ITVal tool: 
The (optimized) upper closure ruleset only contains around 1000 rules and the lower closure only around 500 rules.
Thus, we expected that ITVal could process these rulesets significantly faster.
However, the opposite is the case: ITVal requires more than 10 times the resources (both CPU and RAM, we had to move the analysis to a > \SI{40}{\giga\byte} RAM cluster) to finish the analysis of the closures.
We assume that this is due to the fact that ITVal now understands \emph{all} rules.

\section{Conclusion}
This work was motivated by the fact that we could not find any tool which helped analyzing our lab's and other firewall rulesets.
Though much related work about firewall analysis exists, all academic firewall models are too simplistic to be applicable to those real-world rulesets.
With the transformations presented in this paper, they became processable by existing tools.
With only a small amount of manual inspection, we found previously unknown issues in four real-world firewalls.

We introduced an approximation to reduce even further the complexity of real-world firewalls for subsequent analysis.
In our evaluation, we found that the approximation is good enough to provide meaningful results.
In particular, using further tools, we were finally able to provide our administrator with a meaningful answer to the question of how our firewall partitions the IP space.

Our transformations can be extended for different firewall configurations. 
A user must only provide a primitive matcher for the firewall match conditions she wishes to support.
Since we use ternary logic, a user can specify ``unknown'' as matching outcome, which makes definition of new primitive matchers very easy.
The resulting firewall ruleset conforms to the simple list model in Boolean logic (\ie the common model found in the literature).

Future work includes increasing the accuracy of the approximation by providing more feature-rich primitive matchers and directly implementing firewall analysis algorithms in Isabelle to formally verify them.
Another planned application is to assist firewall migration between different vendors and migrating legacy firewall systems to new technologies.
In particular, such a migration can be easily prototyped by installing a new firewall in chain with the legacy firewall such that packets need to pass both systems: with the assumption that users only complain if services no longer work, the formal argument in this paper proves that the new firewall with an upper closure ruleset operates without user complaints. 
A new fast firewall with a lower closure ruleset allows bypassing a slow legacy firewall, probably removing a network bottleneck, without security concerns.

\section*{Availability}
\label{sec:availability}
The analyzed firewall rulesets can be found at
\begin{center}
\url{https://github.com/diekmann/net-network}
\end{center}
\noindent
Our Isabelle formalization can be obtained from
\begin{center}
\url{https://github.com/diekmann/Iptables_Semantics}
\end{center}

\section*{Acknowledgments}
A special thanks goes to Andreas Korsten for valuable discussions.  
We thank Julius Michaelis for contributing his Shorewall firewall.
We express our gratitude to both for agreeing to publish their firewalls.
In addition, Julius and Lars Noschinski contributed proofs to the formalization of the IP address space.
Manuel Eberl, Lukas Schwaighofer, and Fabian Immler commented on early drafts of this paper. 
This work was greatly inspired by Tobias Nipkow's and Gerwin Klein's book on semantics in Isabelle~\cite{nipkow2014concretesemantics}.

This work has been supported by the German Federal Ministry of Education and Research (BMBF), EUREKA project SASER, grant 16BP12304, and by the European Commission, FP7 project EINS, grant 288021. 

\theendnotes

{\bibliographystyle{splncs03}
\bibliography{literature}}

\end{document}